\documentclass[11pt,a4paper]{article}

\usepackage[T1]{fontenc}
\usepackage[utf8]{inputenc}
\usepackage[margin=2.5cm]{geometry}
\usepackage{amsmath}
\usepackage{graphicx}
\usepackage{booktabs}
\usepackage[authoryear,round]{natbib}
\usepackage{authblk}

\usepackage{placeins}
\usepackage[hidelinks]{hyperref}

\title{Minimising sample size across analysis strategies in preclinical research: Reduced power or negligible benefit}

\author[1,2]{Nicole Ellenbach
  \thanks{Corresponding author: nellenbach@ibe.med.uni-muenchen.de}}

\author[1]{Juliane C. Wilcke}
\author[1]{Maurizio Morelli}
\author[1]{Elisa~Noltenius}
\author[1]{Simona~Jell}
\author[3,4,5,6]{Sebastian~Kobold}
\author[7,8]{Carolin~Herrmann}
\author[1,2]{Anne-Laure~Boulesteix}

\affil[1]{Institute for Medical Information Processing,
Biometry, and Epidemiology (IBE), LMU Medizin,
Ludwig-Maximilians-Universität München, Munich, Germany}

\affil[2]{Munich Center for Machine Learning (MCML),
Munich, Germany}

\affil[3]{Institute of Clinical Pharmacology,
LMU University Hospital, LMU Medizin,
Ludwig-Maximilians-Universität München, Munich, Germany}

\affil[4]{German Cancer Consortium (DKTK), Partner Site Munich,
DKFZ Heidelberg and LMU University Hospital, LMU Medizin,
Ludwig-Maximilians-Universität München, Germany}

\affil[5]{Einheit für Klinische Pharmakologie (EKLiP),
Helmholtz Zentrum München - German Research Center
for Environmental Health, Neuherberg, Germany}

\affil[6]{German Center for Lung Research (DZL),
partner site Munich, Germany}

\affil[7]{Mathematical Institute,
Heinrich Heine University Düsseldorf,
Düsseldorf, Germany}

\affil[8]{Center for Digital Medicine,
Heinrich Heine University Düsseldorf,
Düsseldorf, Germany}

\date{}

\begin{document}

\maketitle

\begin{abstract}
\noindent Preclinical animal studies often allow multiple valid statistical analysis strategies, which can result in different sample sizes required to achieve a target power. Because reducing animal use is an ethical, financial, and practical priority, researchers have a clear incentive to choose the strategy requiring the fewest animals. This may, however, favour an overly optimistic estimate of the sample size required, particularly when prior information is uncertain or biased. This study investigates how minimising sample size across analysis strategies affects statistical power and animal use. We conceptualise this practice as sample size hacking, describe its underlying mechanism, and evaluate its consequences using a simulation study based on a real preclinical cancer study. We varied effect sizes, the level of uncertainty in prior information, and the degree of follow-up bias to assess the resulting sample size reductions and achieved statistical power. Sample size minimisation across analysis strategies substantially reduced the achieved statistical power in several scenarios, particularly for moderate effect sizes and greater uncertainty in prior information. For larger effect sizes or in the presence of follow-up bias, the impact on power was less pronounced, but the resulting reduction in sample size was also smaller. Selecting an analysis strategy solely because it yields the smallest calculated sample size may in fact compromise statistical power, while offering only limited animal savings in other settings. These findings underscore the importance of justifying and pre-specifying the analysis strategy before sample size planning to limit potentially problematic analytical flexibility, reduce the risk of underpowered studies, and promote transparency in preclinical research.
\end{abstract}


\maketitle

\section{Introduction}
In recent years, there has been growing awareness that data analysts face numerous uncertain choices when analysing empirical data \citep{hoffmann2021multiplicity}. These so-called \emph{researcher degrees of freedom} \citep{Simmons2011FalsePositive} can give rise to a wide range of possible analysis strategies. When such flexibility is combined with selective reporting based on the observed results, it can lead to overly optimistic findings in the literature that are difficult or impossible to replicate \citep{hoffmann2021multiplicity, ioannidis2005most}. To address these concerns, various measures have been suggested, including improving statistical planning and study design through power analyses and the pre-registration of studies, with the aim of fixing key decisions in advance and thereby limiting researcher degrees of freedom during data analysis, as well as reducing opportunities to favour positive results through questionable research practices \citep{munafo2017manifesto,nosek2018preregistration}. 

While we agree that appropriate power analyses and sample size calculations are essential to credible science, we emphasise that sample size planning is itself not free of researcher degrees of freedom. Anyone with experience in sample size planning will recognise that this process involves a certain degree of flexibility. For example, sample size planning often involves flexibility in the choice of the procedure used to determine sample size, such as simulations or one of several sample size formulas. In addition, there may already be flexibility in the choice of the statistical analysis strategy before sample size planning, persisting during this process: for a given research question, several statistical methods (e.g., statistical models or tests) may be considered appropriate (possibly with different specifications concerning, e.g., measurement time points), each relying on different assumptions and leading to different sample size requirements. As these assumptions are difficult to verify fully---particularly during the planning phase---the choice of method cannot be uniquely determined, thus constituting one source of researcher degrees of freedom already before data analysis.

One area in which this issue is particularly pronounced is preclinical animal research. Given the complexity of biological systems and the limited understanding of their underlying mechanisms, researchers often face greater uncertainty when selecting a statistical analysis strategy (especially in exploratory settings, but also to some extent in confirmatory research): multiple plausible variants of the estimand are often available, and prior knowledge is often insufficient. Nevertheless, the analysis strategy must be specified before the main study begins to enable an appropriate sample size calculation; in preclinical animal research, this is typically required for ethical approval of animal experiments \citep{ARRIVEguideline2.0}. Because the main study is designed to be analysed according to the pre-specified strategy and conducted with the corresponding sample size, the researcher degrees of freedom primarily arise during the planning stage rather than the analysis stage.

At the same time, there is intense pressure to minimise sample sizes \citep{bonapersona2021increasing}, often even below $10$ animals per group, due to ethical, financial, and practical reasons. Consequently, determining a sample size that achieves adequate statistical power while avoiding unnecessary use of animals is a key consideration in preclinical animal research. Against this background, it may appear reasonable to calculate the required sample size for each plausible analysis strategy and to select the strategy that yields the smallest sample size while achieving the desired statistical power. At first glance, such an approach seems sensible and unproblematic, as it reconciles ethical imperatives of animal reduction while still meeting formal power requirements for the selected analysis method \citep{Guhad2005, Strech2019_3Rs}.

The information required for sample size planning is often based on prior information such as published studies, pilot experiments, previous experience with similar studies, or expert judgement. However, at the preclinical stage, biological knowledge is often not sufficiently developed to define key quantities for sample size planning reliably, such as what magnitude of an effect would constitute a meaningful change in the outcome. Other required inputs, including the outcome variance or correlations between repeated measurements, are often even more difficult to specify reliably. Although some authors argue that relatively small preliminary datasets should generally not be used for sample size planning \citep{AlbersLakens2018, Bird2025PilotStudies, Kraemer2006}, in practice, it may often appear as the only feasible option when prior knowledge is limited. Using such preliminary information for sample size planning is at least more study-specific than merely relying on lab-specific default sample sizes that are commonly accepted by the relevant ethics committee or on generic assumptions lacking a study-specific justification, such as \lq\lq the clinically relevant difference is equal to the within-group standard deviation''. 

Importantly, relying on preliminary information can introduce several types of bias in sample size estimation. One of them is the so-called follow-up bias, which arises when researchers preferentially pursue follow-up studies if preliminary results suggest that adequate statistical power can be achieved with a manageable sample size. This selection process can lead to systematically inflated effect size estimates and, consequently, underpowered follow-up studies \citep{AlbersLakens2018}. In addition, publication bias and general uncertainty in the information required for sample size calculations, such as effect sizes and variance parameters based on previous experience, can further distort sample size planning \citep{Anderson2017SampleSizePlanning}. To address these issues, a range of mitigation approaches has been proposed, including adjustments of effect size estimates to account for bias and uncertainty \citep{Anderson2017SampleSizePlanning, Safeguard_Power}, and approaches that formally incorporate prior information within a Bayesian framework \citep{Mutze2017}. 
We do not focus on such bias mitigation approaches, as they ignore uncertainty arising from methodological choice.

In this paper, we formalize and investigate a potential bias mechanism that may lead to overly small sample sizes, even in the absence of publication or follow-up bias. Rather than focusing on the uncertainty inherent in preliminary data itself, we examine the consequences of combining this uncertainty with the selection of the analysis strategy that yields the smallest estimated sample size $n$. This analysis strategy may thus be selected because favourable random fluctuations in the observed data (from the literature, pilot experiments, or other previous experience) used to estimate the input parameters make it appear preferable in terms of required sample size, but this may not hold for the true distribution underlying the data to be collected. From a theoretical point of view, exploiting these random fluctuations to perform what we call \lq\lq sample size minimisation'' is expected to lead to a power smaller than the specified target power.

This bias mechanism and its impact have to our knowledge never been formally described and quantitatively investigated in the context of practical sample size calculation. Our aim is to examine potential consequences of such sample size minimisation through the selection of the analysis strategy requiring the smallest $n$, and to raise awareness of the resulting potentially substantial reduction in the probability of detecting true effects. Using a real preclinical animal study as a motivating example, we assess how this practice affects statistical power under varying effect sizes, levels of uncertainty in prior information, and follow-up bias. We investigate this practice in preclinical animal research, where planning-stage uncertainties are particularly prevalent and consequential given the ethical considerations involved, while acknowledging that the underlying mechanism may also apply to other research settings.

The remainder of this paper is structured as follows. The Methods section introduces the concept of sample size minimisation through the selection of a statistical analysis strategy as a form of hacking and uses a simple fictitious example to illustrate its potential impact on statistical power. It then describes a motivating preclinical animal study, including the underlying real data, the plausible analysis strategies, and the corresponding sample size formulas. Finally, the simulation design is presented following the ADEMP framework by \citet{Morris2019}. The Results section reports the simulation findings, which are discussed in the Discussion section in terms of their implications for preclinical research, particularly the importance of pre-specifying analysis strategies before sample size planning. The paper concludes by summarising the main findings and recommendations.

\section{Methods}
\label{sec:minimisation}

\subsection{Illustrative example of sample size minimisation}
To motivate the concept of what we refer to as sample size minimisation, consider the following general fictitious example (our real preclinical example follows later in the manuscript). Suppose you are faced with a research question involving ordinal data. In this hypothetical study, the effectiveness of an intervention on an ordinal outcome with five categories is to be investigated by comparing a control group with an intervention group. Your task is to determine an appropriate analysis strategy and to perform sample size planning before any data have been collected.

Assume that a project partner (e.g., a physician or biologist) has  knowledge from previous studies about the distribution of the ordinal outcome in the control group, given by the category probabilities $p_C=(0.10, 0.20, 0.30, 0.25, 0.15)$. The project partner expects the intervention group to show an increase in the probability of the first category and a decrease in the probability of the second, fourth, and fifth categories. They provide the following estimate for the intervention group: $p_I=(0.25, 0.15, 0.30, 0.20, 0.10)$. Such a shift would be considered clinically relevant.

Given this information, several analysis strategies could reasonably be considered for the ordinal outcome \citep{Selman2024OrdinalRCT}. For instance, one might choose a $t$-test, a Wilcoxon--Mann--Whitney (WMW) test, or a proportional odds (PO) model. Since the data have not yet been collected and the project partner's expectations do not clearly indicate which analysis method is most appropriate, it is currently unclear which of the three will ultimately best fit the observed data. At this stage, no single analysis strategy clearly emerges as the obvious choice. Nevertheless, a decision is required in order to conduct sample size planning. One natural idea would therefore be to determine the required sample size for each plausible analysis strategy and then use this information to guide the choice.

To illustrate this approach, we calculate the estimated sample size per group for each of the three plausible analysis methods using (approximate) sample size formulas for ordinal data, assuming a two-sided significance level of $\alpha = 0.05$ and a target power of $1-\beta = 0.8$. The resulting required sample sizes per group are $n_t=308$ for the two-sample $t$-test \citep{Kieser2020}, $n_{\text{WMW}}=344$ for the Wilcoxon--Mann--Whitney test \citep{Zhao2008}, and $n_{\text{PO}}=340$ for the proportional odds model \citep{White2023}. The corresponding sample size formulas are provided in Appendix \ref{app:sample_size_formulas}.

Since the required sample sizes differ between the three analysis methods, a possible general approach, especially when prior knowledge is scarce and assumptions are therefore more uncertain, would be to select the method requiring the weakest assumptions. This criterion, however, may not always yield an unequivocal choice, since the assumptions for the method(s) requiring the weakest assumptions do not necessarily build a subset of the assumptions for the method(s) requiring the strongest assumptions. Furthermore, a method may require strong assumptions that are often unrealistic in practice but be robust in case they are violated, which is typically the case for the $t$-test when $n$ increases. Against this background, another straightforward idea might suggest itself: simply choose the analysis strategy that requires the smallest sample size. Since using a larger sample size to achieve the same statistical power implies greater financial, temporal, and organisational effort and is ethically critical, this decision might appear entirely reasonable. In this example, applying this idea would lead to selecting the $t$-test, as it reaches the desired power with the smallest sample size.

But does this seemingly sensible idea---selecting the analysis strategy that requires the smallest sample size---have any serious drawbacks? It is well-known that sample size planning should generally be conducted cautiously, for example by employing approaches to address biases in sample size estimates derived from preliminary data \citep{AlbersLakens2018, Anderson2017SampleSizePlanning, CocksTorgerson2013}. So the question is whether the practice of \emph{minimising across several analysis strategies} itself introduces a previously unrecognised risk of reduced statistical power.

\subsection{Concept of sample size minimisation across analysis strategies}
Unlike some other practices that reduce the required sample size---such as assuming a too small variance or engaging in \emph{N-hacking}\footnote{\citet{Reinagel2023NHacking} uses the term \emph{N-hacking} to describe repeatedly conducting hypothesis tests as data accumulate and stopping data collection when the results are satisfactory, while reporting only the results obtained at the final sample size. This practice is also commonly referred to as \emph{optional stopping}.}\citep{Reinagel2023NHacking}---that are widely recognised as questionable research practices, sample size minimisation through the choice of analysis strategy may appear less problematic at first glance. In the following, we explain why this approach can lead to insufficient statistical power, on average, even when the sample size calculated for each candidate strategy is based on the desired power level.

Returning to the fictitious example of an ordinal endpoint, the probability distribution for the intervention group provided by the project partner is clearly subject to uncertainty. In reality, there exists a true underlying probability distribution for the five categories, while the probabilities supplied by the project partner represent only an estimate of this distribution. The same consideration applies to the control group, although the associated uncertainty may be smaller if more information is available. A similar issue arises when sample size planning is based on prior information such as data from earlier studies or pilot experiments. In such cases, the available data represent only a random or perhaps even very selective sample from the underlying population, and the parameters estimated from these data---and subsequently used for sample size calculations---are themselves random variables \citep{Anderson2017SampleSizePlanning}. Although the true population parameters are unknown in practice, we assume for this illustrative example that the true intervention effect is spread more evenly across categories, resulting in the following true probability distributions for the control and the intervention group:
$p_C^{true}=(0.11, 0.19, 0.29, 0.26, 0.15)$ and
$p_I^{true}=(0.16, 0.24, 0.29, 0.19, 0.12)$.
If we now determine the sample sizes required by each of the three analysis methods using the three sample size formulas and the true probability distributions $p_C^{true}$ and $p_I^{true}$ instead of $p_C$ and $p_I$, we obtain 
$n_t=604$ for the $t$-test, 
$n_{\text{WMW}}=586$ for the Wilcoxon--Mann--Whitney test, and 
$n_{\text{PO}}=578$ for the proportional odds model. These values represent the reference sample sizes required for each method to achieve a target power of 0.8. In this case, the proportional odds model yields the smallest sample size estimate, rather than the $t$-test, which would have been selected earlier under the sample size minimisation approach.

More generally, when sample size calculations are based on estimated planning pa\-ram\-e\-ters---whether obtained from expert assessments or from prior data---the resulting sample sizes will typically not provide the desired power under the true population parameters. Depending on whether the planning parameters (e.g., the parameters of the assumed probability distributions in the ordinal example) overestimate or underestimate the true population parameters, the calculated sample sizes may be larger or smaller than those required under the true population parameters. Consequently, the true power corresponding to the calculated sample size may be higher or lower than the target power. In fact, the average true power across possible parameter estimates tends to be lower than the target power because of the non-linear relationship between sample size and power \citep{AlbersLakens2018}. The amount of this underpowering varies depending on the accuracy of the parameter estimates, where less accurate estimates (or smaller prior datasets) lead to more underpowering, on average \citep{Bird2025PilotStudies}.

The point we make in the current paper is that this situation gets even worse for sample size minimisation across several candidate analysis strategies. For a given set of parameter estimates, the random deviations from the true population parameters may affect different candidate analysis strategies to varying degrees. As a result, the sample sizes calculated from this set of parameter estimates will typically differ from their respective true required sample sizes across analysis strategies. For some strategies, the calculated sample sizes may be close to the true required sample sizes, whereas for others they may be substantially smaller or larger. If the analysis strategy requiring the smallest sample size is then selected, this effectively means choosing the most favourable alternative from the researcher's practical or ethical perspective among several sample size estimates that are all subject to random variability. However, this comes at a price. The selected sample size will tend to fall in the lower tail of the distribution of possible sample size estimates; in other words, selecting the analysis strategy with the smallest calculated sample size effectively means selecting the most extreme downward fluctuation among several random sample size estimates, and consequently the most extreme downward fluctuation in the implied statistical power. The resulting power under the true population parameters will therefore, on average, be subject to an even more pronounced systematic bias than without sample size minimisation: it will be lower than the corresponding power for each candidate analysis strategy and thus even further below the target power. 

Sample size minimisation through the choice of analysis strategy should therefore not be regarded as a researcher degree of freedom that can be exploited without further consideration. Rather, it represents a questionable research practice and may appropriately be described as \emph{sample size hacking}, in analogy to terms such as \emph{p-hacking} \citep{Simmons2011FalsePositive}, \emph{C-hacking} \citep{Sonabend2022Chacking}, or \emph{N-hacking} \citep{Reinagel2023NHacking}. Following these theoretical considerations, we turn to a practical motivating study based on real preclinical animal data.

\subsection{Motivating preclinical study}
\label{Real_Datasets}
We now consider a preclinical animal experiment that motivates the investigation of sample size minimisation through the choice of analysis strategy. Specifically, we examine a realistic scenario from preclinical cancer research based on actual experimental data, representing a typical in-vivo study conducted in mice to evaluate the efficacy of a novel treatment. In this setting, mice are first injected with tumour cells and, after an initial tumour growth period, randomly assigned to one of two groups: a treatment group receiving the treatment under investigation and a control group receiving a control treatment. Tumour size is measured longitudinally, and survival time is recorded for each animal.

Given the limited prior knowledge about the new treatment---particularly under in-vivo conditions---sample size planning is based on three previous exploratory studies from \citet{Cadilha2021CCR8}. Across these studies, tumour size (in $\text{mm}^2$) was assessed every 2–3 days (up to day 41) in a total of $n = 15$ mice per group (5 per group in each study). 

Such experiments typically involve multiple research questions and potential endpoints \citep{Guo2019MCT}. However, consistent with common practice, we assume that sample size determination is driven by a single primary hypothesis. In the present example, neither the available pilot data nor the underlying biological rationale clearly favours a specific modelling approach. Several outcome definitions are conceivable, including tumour size at a fixed time point and longitudinal tumour progression. Each of these endpoints can be analysed using a range of statistical modelling approaches. For instance, longitudinal tumour measurements could be analysed using linear mixed-effects models and tumour size at a given time point may be evaluated using parametric or non-parametric two-sample tests, with the choice of time point introducing an additional degree of freedom. 

Ideally, sample size planning should be aligned as closely as possible with the intended analysis strategy, which in turn implies different planning approaches. The following section outlines the sample size determination strategies considered for this preclinical study in our simulations.

\subsection{Sample size calculation approaches}
\label{Approaches}
For the simulation study, we focus on tumour size at three selected time points as well as on tumour growth over time. Group comparisons at each time point are performed using a $t$-test and a Wilcoxon--Mann--Whitney test, while the overall group effect across time is analysed using a linear mixed-effects model.

A variety of approaches are available for determining the required sample size for these analyses. To ensure that the computational burden of the simulation study remains manageable, we restrict attention to formula-based methods and do not consider simulation-based sample size calculations. Moreover, for each analysis method, we select a single representative formula from the available options. This choice reflects our primary objective, which is not to compare different sample size determination methods, but rather to investigate the consequences of sample size minimisation.

The corresponding sample size formulas and the respective power formulas for the considered analysis methods are given as follows. All sample size formulas refer to the required sample size per group under balanced group sizes, with all tests conducted two-sided. In all formulas, $z_q$ denotes the $q$-quantile of the standard normal distribution.

For the two-sample $t$-test, we compute the required sample size with the classical formulation of power for the $t$-test under the noncentral t-distribution \citep[e.g.][]{Cohen1988}, as implemented in the R-function \texttt{power.t.test} from the \texttt{stats} package. 
For a given per-group sample size $n$, the power is defined as
\begin{equation*}
1 - \beta = P\bigl( T_{\nu,\delta_n} > t_{1-\alpha/2,\nu} \bigr),
\end{equation*}
where $T_{\nu,\delta_n}$ denotes a noncentral t-distributed random variable with degrees of freedom $\nu = 2n - 2$  and noncentrality parameter 
\begin{equation*}
\delta_n = \frac{\delta}{\sigma} \sqrt{\frac{n}{2}},
\end{equation*}
with $\delta = \mu_C - \mu_I$ denoting the difference in group means and $\sigma$ the common (pooled) standard deviation. $t_{1-\alpha/2,\nu}$ denotes the $(1-\alpha/2)$-quantile of the central t-distribution with $\nu$ degrees of freedom. The required sample size is then obtained numerically as the smallest integer $n$ such that the target power level $1-\beta$ is achieved.

For the Wilcoxon--Mann--Whitney test, we use the Noether approximation \citep{Noether1987} for continuous data without ties. For balanced designs, the required sample size per group is given by
\begin{equation*}
n_{\text{WMW}} = \left\lceil
\frac{(z_{1-\alpha/2}+z_{1-\beta})^2}
{6 (p - \tfrac12)^2}
\right\rceil,
\end{equation*}
where $p=P(Y_C > Y_I)+0.5P(Y_C = Y_I)$ denotes the relative effect, with values greater than 0.5 indicating smaller tumour sizes in the treatment group. The corresponding power for a given per-group sample size $n$ is
\begin{equation*}
1-\beta = \Phi\!\left(
\sqrt{6n (p - \tfrac12)^2} - z_{1-\alpha/2}
\right),
\end{equation*}
where $\Phi(\cdot)$ denotes the standard normal cumulative distribution function.

For the linear mixed-effects model, we use the sample size formula of \citet{ArdEdland2011}, as implemented in the R package \texttt{longpower} \citep{longpower}, for testing the group-by-time interaction under a linear random slope model. The required sample size per group is given by
\begin{equation*}
n_{\text{LMM}}
= \left\lceil\frac{2 (z_{1-\alpha/2} + z_{1-\beta})^2}{\delta^2}\left(\sigma_{s}^2 + \frac{\sigma_e^2}{\sum_j (t_j - \bar{t})^2}\right)\right\rceil,
\end{equation*}
where $\delta$ denotes the fixed effect of the group-by-time interaction, $t_j$ the measurement times ($j=1,\dots,J$), $\bar t$ their mean, $\sigma_s^2$ the variance of the random slope, and $\sigma_e^2$ the residual variance. The corresponding power for a given sample size $n$ is
\begin{equation*}
1-\beta = \Phi\!\left(
\sqrt{
\frac{n \,\delta^2}{
2 \left(
\sigma_{s}^2
+
\frac{\sigma_e^2}{\sum_j (t_j - \bar{t})^2}
\right)
}
}
-
z_{1-\alpha/2}
\right).
\end{equation*}

\subsection{Simulation design}
\label{sec:simulation}
\subsubsection{Aims of the simulation study}
The aim of this simulation study is to investigate how selecting the analysis strategy based on the minimum required sample size affects the resulting true statistical power under different conditions. In particular, we examine how this effect depends on the magnitude of the true effect size and on the presence and extent of follow-up bias, defined here as the selective continuation of studies only if the required sample size remains below a specific feasibility threshold. In addition, we assess how varying levels of uncertainty in the parameters used for sample size planning---reflected by different sizes of preliminary datasets---impact the observed behaviour.

\subsubsection{Data-generating mechanism}
We used the three real preliminary datasets from the preclinical animal trial on a new cancer treatment described earlier as the basis for data generation. To this end, a joint model including a longitudinal submodel and a survival submodel was fitted to the combined data from all three datasets, with time measured from the day of injection of the treatment or control substance. With $Y_{ij}$ denoting the log-transformed tumour size for subject $i$ at observation time $t_{ij}$, for $i = 1, \dots, N$ and $j = 1, \dots, n_i$ and $\text{Group}_i$ denoting a binary indicator for treatment group, the longitudinal submodel was modelled using a linear mixed-effects model of the form
\begin{equation*}
Y_{ij} = m_i(t_{ij}) + \varepsilon_{ij},
\end{equation*}
where the subject-specific expected value is given by
\begin{equation*}
m_i(t) = \beta_0 + \beta_1 \text{Group}_i + \beta_2 t + \beta_3 (\text{Group}_i \cdot t) + b_{1i} t.
\end{equation*}
Here, $b_{1i} \sim \mathsf{N}(0, \sigma_b^2)$ denotes a subject-specific random slope for time
and $\varepsilon_{ij} \sim \mathsf{N}(0, \sigma^2)$ are independent and identically distributed measurement errors.
The random effects $b_{1i}$ and the errors $\varepsilon_{ij}$ are assumed to be mutually independent. No random intercept was specified to reflect the fact that tumour sizes were very similar across animals at the start of treatment. The survival submodel was modelled using a proportional hazards model with a Weibull baseline hazard: 
\begin{equation*}
h_i(t) = h_0(t)\,\exp\big( \gamma_0 + \gamma_1 \,\text{Group}_i + \alpha \, m_i(t) \big),
\end{equation*}
where $h_i(t)$ denotes the hazard function for subject $i$ and the baseline hazard function is given by
\begin{equation*}
h_0(t) = \lambda \rho t^{\rho - 1}.
\end{equation*}
Here, $\lambda > 0$ and $\rho > 0$ denote the scale and shape parameters of the Weibull distribution, respectively. Furthermore, $\gamma_0$ is the intercept of the survival linear predictor (acting as a multiplicative scaling factor on the hazard), $\gamma_1$ is the regression coefficient for the treatment group, and $\alpha$ is the association parameter linking the longitudinal and survival submodels such that the hazard depends on the subject-specific expected tumour size at time $t$, as obtained from the longitudinal submodel. 

For the purpose of data generation, selected parameters of the fitted model were slightly modified in order to obtain simulated tumour growth trajectories that more closely resemble the observed ones and to make the generated datasets more suitable for the simulation. In particular, the treatment main effect $\beta_1$ was reduced, reflecting the assumption that the treatment effect primarily manifests over time rather than through baseline differences between groups. Accordingly, a large effect $\beta_3=-0.04$ and a moderate effect $\beta_3=-0.02$ were set for the treatment $\times$ time interaction. In addition, the residual standard deviation $\sigma$ was reduced to decrease within-subject variability, and the intercept of the survival linear predictor $\gamma_0$ (which acts as a multiplicative scaling factor on the baseline hazard) was adjusted to ensure that a sufficient number of animals remained alive at the relevant analysis time points. Table \ref{tab:parameter_modifications} in Appendix \ref{app:joint_model_parameters} compares the originally fitted and modified values of the four model parameters.

Based on this modified joint model, data were generated using a slightly adapted version of the \texttt{simulate} function from the R package \texttt{JM} \citep{rizopoulos2010jm} for daily tumour sizes until the censoring time of 50 days. The function was modified to ensure equal group sizes for treatment and control, and to draw tumour sizes from a truncated normal distribution in order to avoid negative values.

To reflect different magnitudes of the treatment effect, two scenarios were considered, corresponding to a moderate and a larger effect. For each of these scenarios, two types of datasets were generated. First, a very large dataset comprising 100,000 animals (50,000 per group) was generated to approximate the underlying population distribution. This dataset was used to determine the ``true'' required sample sizes for the different analysis methods and to evaluate the ``true'' statistical power for different sample sizes. Second, for each scenario, 10,000 smaller datasets were generated with sample sizes of $n_{prelim} \in \{5,10,15,20\}$ per group. These datasets represent preliminary data used for sample size planning. The variation in sample size serves as a proxy for different levels of uncertainty in the parameter estimates, with smaller datasets corresponding to higher uncertainty. At the same time, this setup can also be interpreted as reflecting different levels of uncertainty in expert-based parameter specifications.

\subsubsection{Estimands and other targets}
For each simulation run, the quantities of interest are the sample sizes obtained from the respective sample size calculation procedures for each analysis strategy, including the sample size minimisation approach, as well as the corresponding true statistical power under the underlying population model. Here, the term ``true statistical power'' refers to the power calculated using the parameters of the large dataset representing the underlying population distribution.

\subsubsection{Statistical methods}
We compared the calculated sample sizes and corresponding true statistical powers of seven candidate analysis strategies with those obtained under the sample size minimisation approach. The seven strategies comprised a $t$-test and a Wilcoxon--Mann--Whitney test (WMW test), each evaluated at three different time points after treatment ($t_1=10$, $t_2=14$, $t_3=17$), as well as a linear mixed model (LMM) on daily tumour sizes. All strategies were considered in their two-sided form, with a significance level of 0.05 and a target power of 0.8.

Sample sizes for the $t$-test were calculated using the R function \texttt{power.t.test}, based on the difference in group means and the pooled standard deviation estimated from the respective dataset. For the WMW test, the relative effect was estimated using the test statistic returned by the \texttt{wilcox.test} function, divided by the product of the group sizes. This estimate was then used in a slightly modified version of the \texttt{noether} function from the \texttt{RankFD} package, where the restriction that probabilities cannot take values of 0 or 1 was removed. For the linear mixed model, a model with fixed effects for group, time, and their interaction, as well as a random slope for time, was fitted using the \texttt{lme} function. Consistent with the joint model used for data generation, no random intercept was included, reflecting the assumption of similar tumour sizes at the start of treatment. To estimate the required sample size, the interaction effect, the residual variance, and the random slope variance were extracted from the fitted model and used, together with the measurement times, in the \texttt{edland.linear.power} from the \texttt{longpower} package, which implements analytical sample size calculations for longitudinal linear mixed-effects models targeting differences in slopes between groups following \citet{ArdEdland2011}. 

In a first step, these procedures were applied to the large dataset approximating the underlying population distribution in order to determine the ``true'' required sample sizes for all seven analysis strategies under both treatment effect scenarios.

In a second step, for each combination of treatment effect scenario and preliminary sample size ($n_{prelim}\in\{5,10,15,20\}$), sample sizes were calculated for each of the 10,000 simulated preliminary datasets. For each method, the ``true'' statistical power corresponding to the estimated sample size was then evaluated by inserting this estimated sample size into the respective power formula, while using the remaining parameters obtained from the large population dataset. Specifically, power was calculated as described earlier: for the $t$-test using the noncentral t-distribution as implemented in \texttt{power.t.test}, for the Wilcoxon--Mann--Whitney test using the analytical formula based on the Noether approach, and for the linear mixed model according to \citet{ArdEdland2011}, as implemented in \texttt{edland.linear.power}.

For sample size minimisation, the smallest sample size across all seven candidate analysis strategies was selected for each iteration, and the corresponding true power was recorded. In cases where multiple strategies yielded the same minimal sample size, the true power was defined as the mean of the corresponding powers.

To investigate the impact of follow-up bias, a selection mechanism reflecting feasibility constraints was introduced. For each preliminary dataset, sample sizes and corresponding true powers were first computed for all candidate analysis strategies as described above. Analysis strategies requiring a sample size exceeding a predefined feasibility threshold were then excluded, thereby reducing the number of datasets on which the results for the respective strategy were based. As long as at least one analysis strategy satisfied the feasibility criterion, sample size minimisation was subsequently performed by selecting the strategy with the smallest sample size among the remaining candidates. The number of datasets underlying the sample size minimisation results decreased only if none of the strategies achieved a sufficiently small sample size, in which case the number of datasets for all other strategies decreased as well.

This procedure, representing a simple form of follow-up bias, was applied using two different feasibility thresholds, $n_{FUB}\in\{25,50\}$, where a smaller threshold corresponds to more restrictive feasibility constraints and thus stronger follow-up bias.

In summary, we compared the sample size minimisation approach with seven fixed analysis strategies across two treatment effect scenarios and four preliminary sample sizes, both in the absence of follow-up bias and under two levels of follow-up bias. All analyses and simulations were conducted using R (version 4.5.0) on Linux. The corresponding R code and data are available at \url{https://github.com/lmu-ibe-bmm/SampleSizeMinimisation}.

\subsubsection{Performance measures}
Performance was assessed by summarising results across simulation runs. Specifically, we evaluated the distribution of the calculated sample sizes and the corresponding true statistical power (as defined above) for each strategy and scenario. In addition, we considered the proportion of preliminary datasets for which the achieved power fell below the target level of 0.8 and, for the scenarios with follow-up bias, the number of preliminary datasets for which the estimated sample size did not exceed the feasibility threshold and was therefore retained in the analysis. We further examined the frequency with which each candidate analysis strategy yielded the smallest sample size under the sample size minimisation approach.

\section{Results}
\label{sec:results}
\begin{figure}
    \centering
    \includegraphics[width=\linewidth]{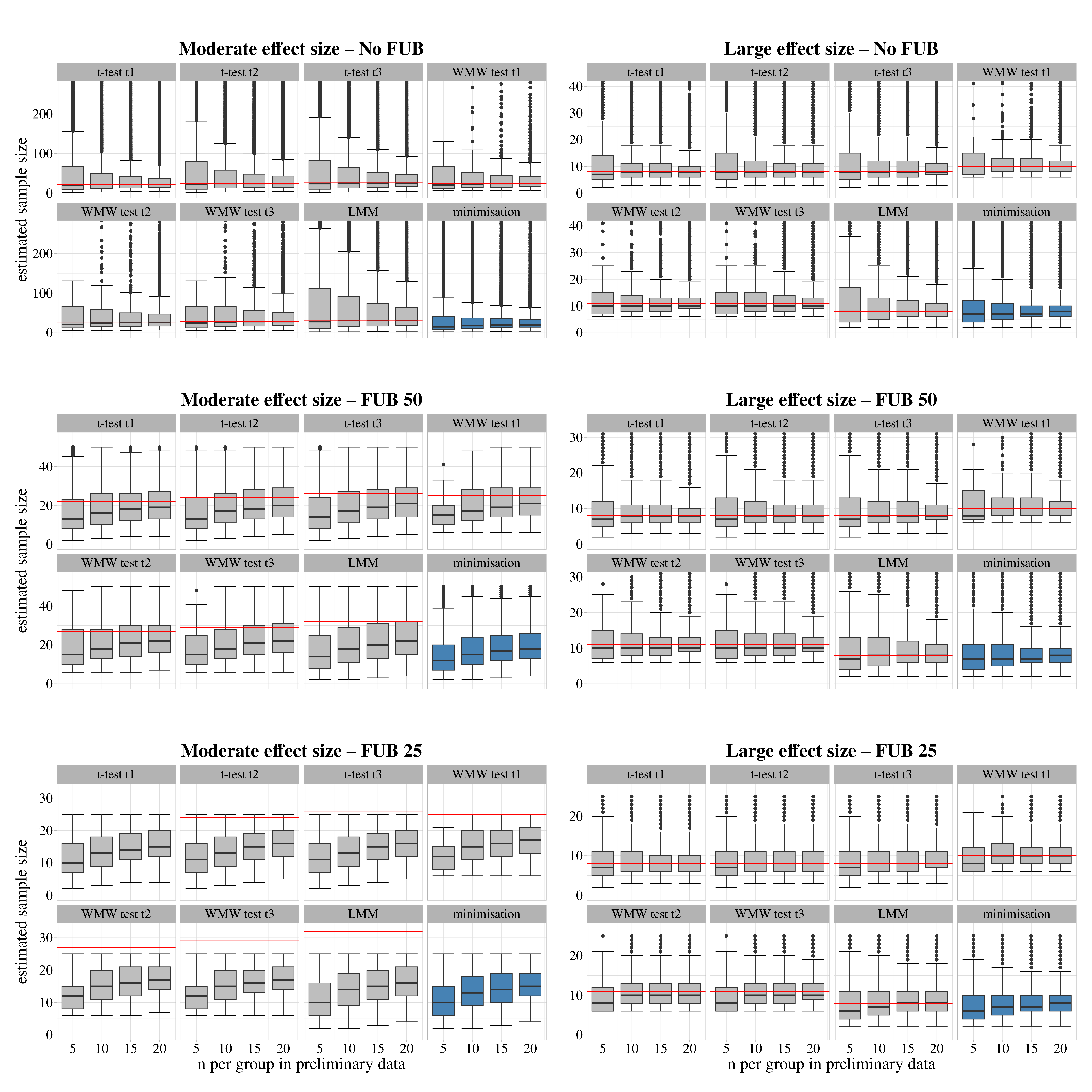}
    \caption{Estimated sample sizes across analysis strategies: Estimated sample sizes per group for each $n_{prelim}$ for the $t$-test, Wilcoxon--Mann--Whitney (WMW) test, linear mixed model (LMM), and sample size minimisation for a moderate (left) and a large (right) treatment effect. The rows correspond to no follow-up bias (top), a follow-up bias threshold of $n_{FUB}=50$ (middle), and a follow-up bias threshold of $n_{FUB}=25$ (bottom). Outliers are displayed; however, the y-axes were restricted to improve the readability of the boxplots in the presence of extreme values. The horizontal lines indicate the true sample sizes (estimated with the parameters from the large dataset) to reach a power of 0.8.}
    \label{fig:boxplot_n}
\end{figure}

The simulation results quantify the impact of sample size minimisation across analysis strategies. Figure \ref{fig:boxplot_n} displays the estimated sample sizes for both effect sizes (left: moderate; right: large) and the three considered follow-up bias scenarios (top: no follow-up bias; middle: follow-up bias with threshold $n_{FUB}=50$; bottom: follow-up bias with threshold $n_{FUB}=25$). For each scenario, boxplots are shown for all considered analysis methods, including sample size minimisation (highlighted in blue), illustrating the distribution of estimated sample sizes across simulation iterations for the four preliminary sample sizes $n_{prelim}\in\{5,10,15,20\}$. Since preliminary datasets with estimated sample sizes exceeding the respective follow-up bias thresholds were excluded, the follow-up bias scenarios are based on fewer preliminary datasets. The exact numbers are reported in Table \ref{tab:remaining_iterations} in Appendix \ref{app:additional_results}.

In the absence of follow-up bias (top row), a clear pattern emerges: the variability of the estimated sample sizes decreases as $n_{prelim}$ increases. This reflects the reduced uncertainty in parameter estimation based on larger preliminary datasets. The reduction in variability is more pronounced for the moderate effect size than for the large effect size. 
For both effect sizes, the median sample sizes are slightly underestimating the true sample sizes at $n_{prelim}=5$ and align well with the true sample sizes for larger preliminary datasets. Without follow-up bias, the linear mixed model yields the largest sample sizes and shows the greatest variability. Compared to all other analysis strategies, sample size minimisation results in smaller sample sizes with reduced variability. The corresponding median sample sizes are provided in Table \ref{tab:median_sample_size} in Appendix \ref{app:additional_results}. This difference is substantially less pronounced in the large effect scenario than for the moderate effect size.

When introducing follow-up bias (middle and bottom rows), a different pattern becomes apparent. For the moderate effect size under feasibility constraints ($n_{FUB}=50$ and $n_{FUB}=25$), the decreasing variability with increasing $n_{prelim}$ is no longer observed. Instead, the interquantile ranges are smallest for $n_{prelim}=5$ and remain relatively similar for larger preliminary sample sizes. This is due to the imposed upper limits, which truncate large sample size estimates and effectively remove extreme overestimations---which occurred more frequently for smaller preliminary group sizes---from the analysis. The median sample sizes for all analysis strategies fall below the true required sample sizes, with stronger underestimation observed for smaller preliminary datasets (see Tables \ref{tab:median_sample_size_followup50} and \ref{tab:median_sample_size_followup25} in Appendix \ref{app:additional_results}). In contrast to the scenario without follow-up bias, no single analysis strategy yields substantially larger sample sizes; instead, the distributions across strategies become more similar. Sample size minimisation, which previously produced distinctly smaller sample sizes, now largely converges with the other analysis strategies.

This effect becomes even more pronounced under the stronger follow-up bias ($n_{FUB}=25$). Here, all median sample sizes lie well below the true value, and for some analysis strategies, the entire boxplot falls below the true required sample size, as this value is no longer attainable under the imposed constraint. In this setting, sample size minimisation is virtually indistinguishable from the other strategies.

For the large effect size, the impact of follow-up bias is considerably smaller. The distributions remain largely similar across all three scenarios, with only a slight reduction in variability and median sample sizes under the strongest constraint ($n_{FUB}=25$), particularly for smaller preliminary datasets. This is because the true required sample sizes are already relatively small for large effects and therefore remain below the feasibility thresholds. In addition, the larger effect is typically already more clearly reflected in the preliminary data, leading to consistently smaller estimated sample sizes across simulation runs. As a result, most of the sample size estimates fall below the imposed thresholds, so that the selection mechanism induced by follow-up bias has little impact on the resulting distributions.

\begin{figure}
    \centering
    \includegraphics[width=\linewidth]{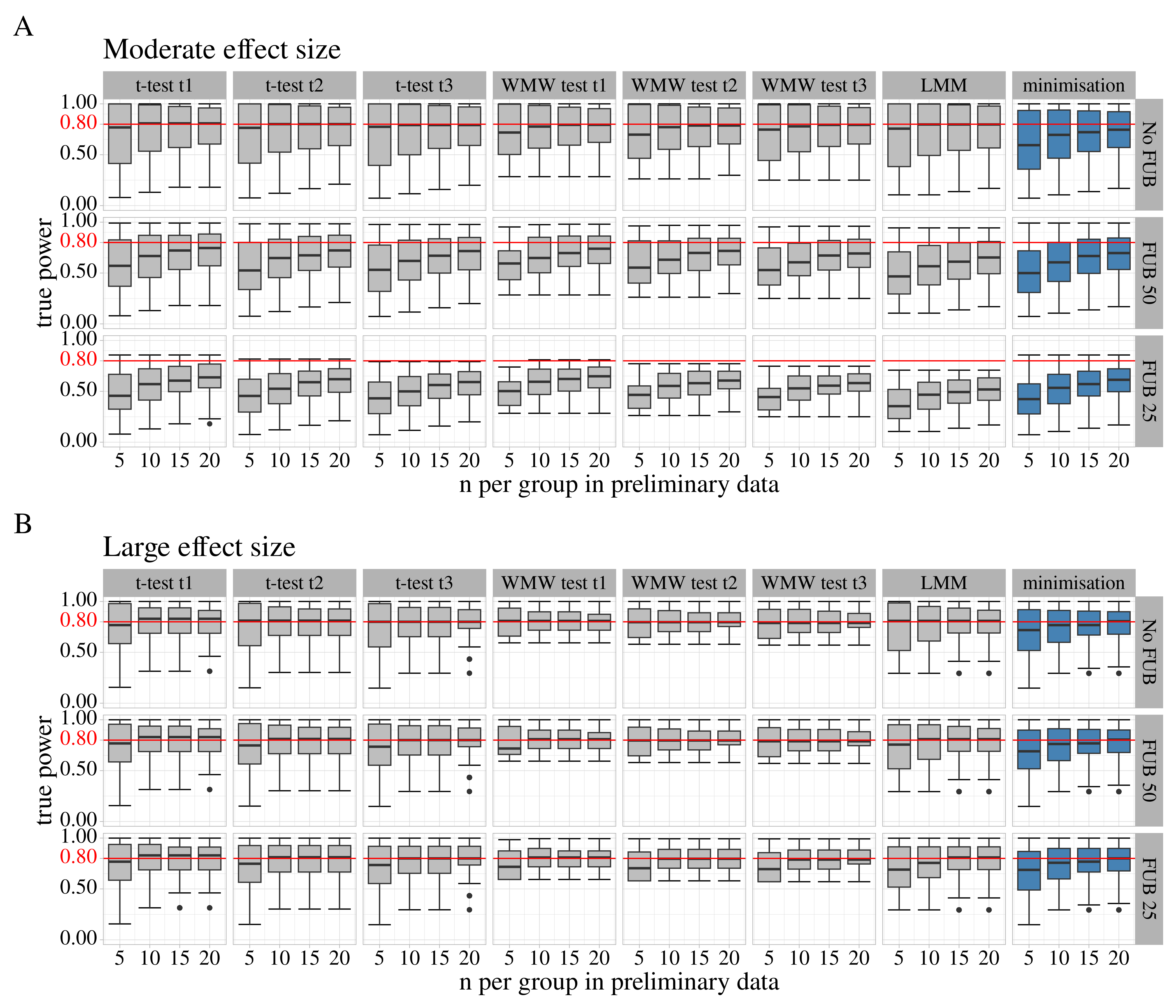}
    \caption{True statistical power across analysis strategies: True power for a moderate (A) and large (B) treatment effect for each $n_{prelim}$ for the $t$-test, Wilcoxon--Mann--Whitney (WMW) test, linear mixed model (LMM), and sample size minimisation. The upper rows show the true power without follow-up bias (No FUB), whereas the middle and lower rows depict two levels of follow-up bias by assuming that only methods with calculated sample sizes below the respective thresholds of $n_{FUB}=50$ (FUB 50) or $n_{FUB}=25$ (FUB 25) would be considered.}
    \label{fig:boxplot_power}
\end{figure}

Figure \ref{fig:boxplot_power} shows the true statistical power corresponding to the estimated sample sizes for the moderate and large effect size scenarios, respectively. The true power was calculated by combining the estimated sample sizes with the true parameters from the underlying population. For both effect sizes, boxplots are presented for each of the three considered follow-up bias scenarios (top: no follow-up bias; middle: follow-up bias with threshold $n_{FUB}=50$; bottom: follow-up bias with threshold $n_{FUB}=25$), for all considered analysis strategies (with sample size minimisation highlighted in blue), and for all four preliminary sample sizes $n_{prelim}\in\{5,10,15,20\}$.

In the absence of follow-up bias (top rows of Figure \ref{fig:boxplot_power}), the variability of the achieved power decreases as the size of the preliminary data increases. In addition, the median power is below the target level of 0.8 for smaller values of $n_{prelim}$. Both effects are less pronounced for the large effect size (part B of Figure \ref{fig:boxplot_power}). Sample size minimisation yields underpowered median values for all considered preliminary sample sizes, except for the largest preliminary sample size in the large effect scenario. Overall, the medians are lower for sample size minimisation, but the range of achieved power is similar for sample size minimisation and the individual analysis strategies.

\begin{table}[ht]
\caption{Fraction of underpowered iterations (power $<$ 0.8) per $n_{prelim}$ and effect size without follow-up bias} 
\label{tab:prob_below0.8}
\begin{tabular*}{\textwidth}{@{\extracolsep{\fill}}l|rrrr|rrrr@{\extracolsep{\fill}}}
 \toprule
 & \multicolumn{4}{c|}{Moderate effect size} & \multicolumn{4}{c}{Large effect size} \\
 \cline{2-5}\cline{6-9}
$n_{prelim}$ & 5 & 10 & 15 & 20 & 5 & 10 & 15 & 20 \\
\midrule
$t$-test t1 & 0.513 & 0.500 & 0.497 & 0.493 & 0.502 & 0.469 & 0.463 & 0.451 \\ 
  $t$-test t2 & 0.512 & 0.499 & 0.499 & 0.494 & 0.476 & 0.446 & 0.431 & 0.407 \\ 
  $t$-test t3 & 0.520 & 0.509 & 0.506 & 0.502 & 0.469 & 0.427 & 0.406 & 0.382 \\ 
  WMW test t1 & 0.531 & 0.517 & 0.507 & 0.511 & 0.471 & 0.462 & 0.451 & 0.433 \\ 
  WMW test t2 & 0.511 & 0.513 & 0.515 & 0.510 & 0.562 & 0.534 & 0.555 & 0.534 \\ 
  WMW test t3 & 0.567 & 0.526 & 0.515 & 0.511 & 0.546 & 0.521 & 0.531 & 0.507 \\ 
  LMM & 0.529 & 0.509 & 0.510 & 0.512 & 0.488 & 0.459 & 0.447 & 0.428 \\ 
  minimisation & 0.642 & 0.605 & 0.586 & 0.574 & 0.588 & 0.531 & 0.514 & 0.494 \\ 
   \bottomrule
\end{tabular*}
\footnotetext{WMW: Wilcoxon--Mann--Whitney; LMM: linear mixed model; minimisation: sample size minimisation across all analysis strategies; $t$-test and WMW at three different time points t1, t2, t3}
\end{table}

Table \ref{tab:prob_below0.8} summarises the probability of underpowering in the scenarios without follow-up bias. Here, the probability of underpowering is estimated as the proportion of simulation iterations with an achieved true power below 0.8. Across methods, the probability of underpowering is generally close to 0.5, although it tends to be slightly higher for smaller preliminary sample sizes and slightly lower for larger preliminary sample sizes. The larger effect size is associated with somewhat lower probabilities of underpowering, particularly for the $t$-test at the second and third time points, the Wilcoxon--Mann--Whitney test at the first time point, and the linear mixed model, where the probabilities remain below 0.5 across all considered values of $n_{prelim}$. For the moderate effect size, sample size minimisation consistently results in higher probabilities of underpowering than all individual analysis strategies. Notably, even the lowest probability observed for sample size minimisation ($n_{prelim}=20$) exceeds the highest probability observed for any individual strategy (WMW test at time point 3 with $n_{prelim}=5$). In contrast, for the large effect size, sample size minimisation yields a clearly elevated underpowering probability only for the smallest preliminary sample size ($n_{prelim}=5$). For larger preliminary sample sizes, the underpowering probability under sample size minimisation approaches 0.5 and becomes comparable to, or even slightly lower than, the corresponding probabilities of some individual strategies.

Under follow-up bias, a different pattern emerges for the moderate effect size. Compared with the scenario without follow-up bias, the reduction in variability with increasing $n_{prelim}$ becomes less pronounced, while the interquartile ranges decrease as the follow-up bias becomes stronger. The true power becomes increasingly underestimated for smaller preliminary sample sizes and for stronger follow-up bias. With a threshold of $n_{FUB}=50$, the median power for all methods and all preliminary sample sizes already falls below the target level of 0.8. Under the stronger follow-up bias ($n_{FUB}=25$), all boxes---and for some analysis strategies even the entire ranges---lie below a power level of 0.8. In this setting, sample size minimisation no longer differs noticeably from the individual strategies, neither with respect to the distribution nor to the median achieved power.

For the large effect size, in contrast, the effect of follow-up bias on true power is minimal. The distributions remain largely unchanged across the three follow-up bias scenarios, with only a slight decrease in median power observed for some individual methods for smaller preliminary sample sizes as follow-up bias becomes stronger. The median power under sample size minimisation remains relatively stable across the different follow-up bias scenarios. Consequently, the negative effect of sample size minimisation for smaller preliminary sample sizes becomes less pronounced, as the achieved power of several individual strategies also shifts towards similarly underpowered values.

These effects of follow-up bias are likewise reflected in the proportion of underpowered iterations, as shown in Tables \ref{tab:prob_below0.8_followup50} and \ref{tab:prob_below0.8_followup25} in Appendix \ref{app:additional_results} for $n_{FUB}=50$ and $n_{FUB}=25$, respectively. 

Table \ref{tab:minfreq_all} in Appendix \ref{app:additional_results} shows for all considered scenarios the percentage of preliminary datasets for which each strategy yielded the smallest estimated sample size used for sample size minimisation. Across all considered scenarios, every analysis strategy yielded the smallest sample size for at least some preliminary datasets; however, the frequency with which this occurred differed between analysis strategies. For a moderate effect size, the $t$-test at the first time point most frequently resulted in the smallest sample size. For a large effect size, the smallest sample sizes were most frequently obtained using either the $t$-test at the first time point or the linear mixed model.

\section{Discussion}
\label{sec:discussion}

In this paper, we characterised how minimising sample size across multiple candidate analysis strategies, driven by the ethical and practical pressure to reduce animal use, can reduce the statistical power of preclinical studies. Our simulation results suggest that, particularly in scenarios with moderate effect sizes and small preliminary datasets, the reduction in sample size resulting from minimisation was accompanied by a clear loss in statistical power. By contrast, in scenarios with large effect sizes or in the presence of follow-up bias, the negative impact of sample size minimisation on power became much less pronounced. However, in these latter scenarios, sample size minimisation also offered little practical advantage, because the resulting sample sizes were no longer meaningfully smaller than those obtained from the individual analysis strategies. In summary, these findings suggest that sample size minimisation either increases the risk of underpowered studies or provides little practical benefit when this risk is attenuated. These findings therefore support treating the selection of an analysis strategy as a methodological decision to be made, justified, and pre-specified before sample size planning. 

The consequences of minimising sample size across analysis strategies differ from those of p-hacking and related practices. P-hacking can inflate the type I error rate by exploiting analytical flexibility to pursue statistically significant results that may benefit authors in the short term (greater chance of publication and attention). Sample size minimisation, by contrast, can reduce statistical power and thereby increase the risk of type II errors. For example, genuine effects may go undetected in replication studies. This can disadvantage study authors and raise ethical concerns regarding animal use in preclinical research. At the same time, smaller sample sizes lead to greater variability in effect estimates, including more extreme estimates, and can contribute to more inconclusive findings due to reduced precision. Selective reporting may then further contribute to the dissemination of more extreme effect estimates and false-positive findings (as the proportion of false positives among statistically significant findings increases with decreasing statistical power \cite{ioannidis2005most}). Thus, sample size minimisation across analysis strategies can be viewed as a form of hacking that, unlike other forms of hacking often investigated in the context of the replication crisis such as p-hacking, primarily increases the risk of type II errors and inconclusive findings, while potentially also contributing to more extreme effect estimates and more false-positive findings being selectively reported. Importantly, this practice need not involve deliberate manipulation: it can arise from the seemingly reasonable objective of reducing animal use.

Our study has several limitations and could be extended in different ways. First, to keep the scope feasible, we focused on theoretical considerations, an illustrative example, and a simulation study based on a single preclinical study. The extent to which sample size minimisation across analysis strategies affects statistical power may vary across applications, but also by which analysis strategies are considered. The choice of analysis strategies inevitably involves some arbitrariness. We focused on the $t$-test and Wilcoxon--Mann--Whitney test at relevant time points and on linear mixed models; other reasonable choices could include repeated-measures ANOVA, Cox or joint models, and different approaches to handling missing data when deriving parameters from the preliminary data. In practice, researchers might also minimise sample size by including or excluding additional covariates within a given analytical method. We did not investigate this possibility separately, but focused on minimising sample size across a limited set of analysis strategies. This restriction was intentional, since minimising sample size across a very large number of strategies is unlikely to reflect typical practice. Future research could investigate how the impact of sample size minimisation varies across different sets of plausible analysis strategies.

Second, in our simulation study, we considered only follow-up bias alongside sample size minimisation, because we anticipated that follow-up bias might modify its impact. We deliberately did not include mitigation approaches for dealing with publication bias or for accounting for uncertainty in parameter estimates derived from prior information, since such approaches are often tailored to specific analysis methods, whereas our investigation of sample size minimisation spans multiple methods. Our goal was to isolate the consequences of sample size minimisation without introducing additional methodological complexities. Future work should examine the interplay between sample size minimisation and these other sources of bias and uncertainty in more detail.

Third, we did not quantitatively investigate the related issue of maximising---based on preliminary data---the power across several analysis strategies while treating sample size as fixed. We expect that random variation in preliminary data, combined with selecting the strategy that yields the maximal apparent power based on those data-derived parameters, will lead to overestimation of the true power, analogous to how minimising sample size across analysis strategies leads to underestimation of the required sample size and thus reduces study power. This mechanism, although very similar to the one described in this paper, warrants specific investigation in future work. 

Our findings reinforce the need for transparent reporting of the rationale for selecting an analysis strategy and the assumptions underlying sample size calculations. Where multiple analysis strategies are scientifically plausible, the choice should be based on scientific and statistical considerations and appropriately justified and specified before sample size planning, with the subsequent sample size calculation aligned with the planned statistical analysis. This is consistent with recommendations proposed for preclinical confirmatory multi-laboratory studies, which emphasise the importance of aligning sample size calculations with the planned statistical analyses \citep{Arroyo-Araujo2026}. Transparently reporting these decisions allows readers to assess the methodological basis of the planned sample size and the resulting statistical power. Importantly, pre-specification needs to be accompanied by adherence to the planned analysis strategy, as discrepancies between study protocols and published reports regarding sample size calculations and statistical analyses have been documented in clinical research \citep{Chan2008}. Such pre-specification helps prevent the selection of an analysis strategy from being influenced by its sample size implications, thereby supporting the reliability and reproducibility of preclinical research. 

\section{Conclusions}
Sample size minimisation across plausible analysis strategies may increase the risk of underpowered preclinical studies, particularly when effect sizes are moderate and preliminary information is limited. In other settings, the reduction in power may be less pronounced, but the practical benefit in terms of animal numbers may also be limited. Importantly, this risk can arise from the seemingly reasonable objective of reducing animal use and does not require deliberate manipulation. Given these negative consequences and the lack of a clear practical benefit in many cases, researchers should select and justify their analysis strategy based on scientific and statistical considerations instead of comparing the sample sizes required by alternative strategies. Transparently documenting these decisions and the assumptions underlying the sample size calculation can help limit analytical flexibility during study planning and support more reproducible preclinical research, while preserving the ethical objective of reducing animal use through scientifically justified study design.

\section*{Conflicts of interest}
S.K. has received honoraria from Plectonic, TCR2 Inc, Regeneron, Miltenyi, Galapagos, Cymab, Novartis, BMS, and GSK. S.K. is an inventor of several patents in the field of immuno-oncology. S.K. received license fees from TCR2 Inc and Carina Biotech. S.K. received research support from TCR2 Inc, Tabby Therapeutics, Catalym, Plectonic, and Arcus Bioscience for work unrelated to the manuscript.
All other authors declare that they have no competing interests.

\section*{Funding}
This work was supported by the German Research Foundation (grants \mbox{BO3139/7-2} and \mbox{BO3131/10-1} to A.L.B.), the German Federal Ministry of Education and Research (to A.L.B. and S.K.), the m4-Award of the Bavarian Ministry for Economic Affairs (to S.K.), and the European Research Council (Starting Grant 756017, PoC Grant 101100460, and CoG 101124203 to S.K.). The authors of this work take full responsibility for its content.

\FloatBarrier

\section*{Data availability}

The data and R code required to reproduce all results are freely available on GitHub: \url{https://github.com/lmu-ibe-bmm/SampleSizeMinimisation}

\section*{Acknowledgements}

Large language models (LLMs) were used to assist with language editing and individual formulations and to support the development of R code for figure generation. The authors reviewed and verified the AI-assisted output and remain responsible for the final manuscript and code.

\FloatBarrier

\bibliographystyle{abbrvnat}
\bibliography{reference_preprint}

@article{AlbersLakens2018,
  author  = {Albers, Casper and Lakens, Dani{\"e}l},
  title   = {When power analyses based on pilot data are biased: inaccurate effect size estimators and follow-up bias},
  journal = {Journal of Experimental Social Psychology},
  year    = {2018},
  volume  = {74},
  pages   = {187--195},
}

@article{Anderson2017SampleSizePlanning,
  author  = {Anderson, Samantha F. and Kelley, Ken and Maxwell, Scott E.},
  title   = {Sample-size planning for more accurate statistical power: a method adjusting sample effect sizes for publication bias and uncertainty},
  journal = {Psychological Science},
  year    = {2017},
  volume  = {28},
  number  = {11},
  pages   = {1547--1562},
}

@article{ArdEdland2011,
  author       = {Ard, Mary C. and Edland, Steven D.},
  title        = {Power calculations for clinical trials in {Alzheimer's} disease},
  journal      = {Journal of Alzheimer's Disease},
  year         = {2011},
  volume       = {26},
  number       = {Suppl 3},
  pages        = {369--377},
  pmid         = {21971476},
  pmcid        = {PMC3568927}
}

@article{Arroyo-Araujo2026,
    author = {Arroyo-Araujo, María and Carneiro, Clarissa F. D. and Piper, Sophie            K. and Wilcke, Juliane C. and Ellenbach, Nicole and Boulesteix, Anne-           Laure and Emprechtinger, Robbert and Ineichen, Benjamin V. and                  Riecken, Lars Bjorn and Haller, Bernhard and Voelkl, Bernhard and               Held, Leonhard and Konietschke, Frank and Toelch, Ulf},
    title = {Data analysis planning and reporting for confirmatory multi-lab preclinical trials: a tutorial},
    journal = {Biometrical Journal},
    volume = {68},
    number = {4},
    pages = {e70152},
    year = {2026}
}

@article{Bird2025PilotStudies,
  author  = {Bird, Alexander D. and Jedlicka, Peter and Wilhelm, Jochen},
  title   = {Animal pilot studies should not be used to estimate sample size if effect size and population variance are unknown},
  journal = {ALTEX},
  year    = {2025},
  volume  = {42},
  number  = {3},
  pages   = {531--535},
  pmid    = {40230102}
}

@article{bonapersona2021increasing,
  title={Increasing the statistical power of animal experiments with historical control data},
  author={Bonapersona, V and Hoijtink, H and RELACS Consortium and Sarabdjitsingh, RA and Jo{\"e}ls, M},
  journal={Nature Neuroscience},
  volume={24},
  number={4},
  pages={470--477},
  year={2021},
  publisher={Nature Publishing Group US New York}
}

@article{Cadilha2021CCR8,
  author = {Cadilha, Bruno L. and Benmebarek, Mohamed-Reda and Dorman, Klara and Oner, Arman and Lorenzini, Theo and Obeck, Hannah and Vänttinen, Mira and Di Pilato, Mauro and Pruessmann, Jasper N. and Stoiber, Stefan and Huynh, Duc and Märkl, Florian and Seifert, Matthias and Manske, Katrin and Suarez-Gosalvez, Javier and Zeng, Yi and Lesch, Stefanie and Karches, Clara H. and Heise, Constanze and Gottschlich, Adrian and Thomas, Moritz and Marr, Carsten and Zhang, Jin and Pandey, Dharmendra and Feuchtinger, Tobias and Subklewe, Marion and Mempel, Thorsten R. and Endres, Stefan and Kobold, Sebastian},
  title = {Combined tumor-directed recruitment and protection from immune suppression enable {CAR T} cell efficacy in solid tumors},
  journal = {Science Advances},
  year = {2021},
  volume = {7},
  number = {24},
  pages = {eabi5781},
}

@article {Chan2008,
	author = {Chan, An-Wen and Hr{\'o}bjartsson, Asbj{\o}rn and J{\o}rgensen, Karsten J and G{\o}tzsche, Peter C and Altman, Douglas G},
	title = {Discrepancies in sample size calculations and data analyses reported in randomised trials: comparison of publications with protocols},
	volume = {337},
	elocation-id = {a2299},
	year = {2008},
	publisher = {BMJ Publishing Group Ltd},
	journal = {BMJ}
}

@article{CocksTorgerson2013,
  author  = {Cocks, K. and Torgerson, D. J.},
  title   = {Sample size calculations for pilot randomized trials: a confidence interval approach},
  journal = {Journal of Clinical Epidemiology},
  year    = {2013},
  volume  = {66},
  number  = {2},
  pages   = {197--201},
}

@book{Cohen1988,
  author    = {Cohen, Jacob},
  title     = {Statistical Power Analysis for the Behavioral Sciences},
  edition   = {2},
  publisher = {Routledge},
  year      = {1988},
}

@article{Guhad2005,
  author  = {Guhad, F.},
  title   = {Introduction to the 3{Rs} (refinement, reduction and replacement)},
  journal = {Journal of the American Association for Laboratory Animal Science},
  year    = {2005},
  volume  = {44},
  number  = {2},
  pages   = {58--59}
}

@article{Guo2019MCT,
  title   = {The design, analysis and application of mouse clinical trials in oncology drug development},
  author  = {Guo, Sheng and Jiang, Xiaoqian and Mao, Binchen and Li, Qi-Xiang},
  journal = {BMC Cancer},
  year    = {2019},
  volume  = {19},
  number  = {718},
  pages   = {1--20},
}

@article{hoffmann2021multiplicity,
  title={The multiplicity of analysis strategies jeopardizes replicability: lessons learned across disciplines},
  author={Hoffmann, Sabine and Sch{\"o}nbrodt, Felix and Elsas, Ralf and Wilson, Rory and Strasser, Ulrich and Boulesteix, Anne-Laure},
  journal={Royal Society Open Science},
  volume={8},
  number={4},
  pages={201925},
  year={2021},
  publisher={The Royal Society}
}

@article{longpower,
  title = {Power and sample size for longitudinal models in {R}-the longpower package and {Shiny} app},
  author = {Samuel Iddi and Michael C. Donohue},
  journal = {R Journal},
  year = {2022},
  volume = {14},
  number = {1},
  pages = {264-281},
  publisher = {R Foundation for Statistical Computing},
}

@article{ioannidis2005most,
  title={Why most published research findings are false},
  author={Ioannidis, John PA},
  journal={PLoS Medicine},
  volume={2},
  number={8},
  pages={e124},
  year={2005},
  publisher={Public Library of Science}
}

@book{Kieser2020,
  author    = {Kieser, Meinhard},
  title     = {Methods and Applications of Sample Size Calculation and Recalculation in Clinical Trials},
  publisher = {Springer},
  series    = {Springer Series in Pharmaceutical Statistics},
  year      = {2020}
}

@article{Kraemer2006,
    author = {Kraemer, Helena Chmura and Mintz, Jim and Noda, Art and Tinklenberg, Jared and Yesavage, Jerome A.},
    title = {Caution Regarding the Use of Pilot Studies to Guide Power Calculations for Study Proposals},
    journal = {Archives of General Psychiatry},
    volume = {63},
    number = {5},
    pages = {484-489},
    year = {2006},
    month = {05},
}

@article{Morris2019,
  author  = {Morris, Tim P. and White, Ian R. and Crowther, Michael J.},
  title   = {Using simulation studies to evaluate statistical methods},
  journal = {Statistics in Medicine},
  year    = {2019},
  volume  = {38},
  number  = {11},
  pages   = {2074--2102}
}

@article{munafo2017manifesto,
  title={A manifesto for reproducible science},
  author={Munaf{\`o}, Marcus R and Nosek, Brian A and Bishop, Dorothy VM and Button, Katherine S and Chambers, Christopher D and Percie du Sert, Nathalie and Simonsohn, Uri and Wagenmakers, Eric-Jan and Ware, Jennifer J and Ioannidis, John PA},
  journal={Nature Human Behaviour},
  volume={1},
  number={1},
  pages={0021},
  year={2017},
  publisher={Nature Publishing Group UK London}
}

@article{Mutze2017,
  author = {M{\"u}tze, Tobias and Schmidli, Heinz and Friede, Tim},
  title = {Sample size re-estimation incorporating prior information on treatment effects},
  journal = {Statistics in Medicine},
  year = {2017},
  volume = {36},
  number = {8},
  pages = {1237--1253},
}

@article{Noether1987,
 author = {Gottfried E. Noether},
 journal = {Journal of the American Statistical Association},
 number = {398},
 pages = {645--647},
 publisher = {[American Statistical Association, Taylor & Francis, Ltd.]},
 title = {Sample Size Determination for Some Common Nonparametric Tests},
 urldate = {2026-05-09},
 volume = {82},
 year = {1987}
}

@article{nosek2018preregistration,
  title={The preregistration revolution},
  author={Nosek, Brian A and Ebersole, Charles R and DeHaven, Alexander C and Mellor, David T},
  journal={Proceedings of the National Academy of Sciences},
  volume={115},
  number={11},
  pages={2600--2606},
  year={2018},
  publisher={National Academy of Sciences}
}

@article{ARRIVEguideline2.0,
    author = {Percie du Sert, Nathalie AND Hurst, Viki AND Ahluwalia, Amrita AND Alam, Sabina AND Avey, Marc T. AND Baker, Monya AND Browne, William J. AND Clark, Alejandra AND Cuthill, Innes C. AND Dirnagl, Ulrich AND Emerson, Michael AND Garner, Paul AND Holgate, Stephen T. AND Howells, David W. AND Karp, Natasha A. AND Lazic, Stanley E. AND Lidster, Katie AND MacCallum, Catriona J. AND Macleod, Malcolm AND Pearl, Esther J. AND Petersen, Ole H. AND Rawle, Frances AND Reynolds, Penny AND Rooney, Kieron AND Sena, Emily S. AND Silberberg, Shai D. AND Steckler, Thomas AND Würbel, Hanno},
    journal = {PLOS Biology},
    publisher = {Public Library of Science},
    title = {The ARRIVE guidelines 2.0: Updated guidelines for reporting animal research},
    year = {2020},
    month = {07},
    volume = {18},
    pages = {1-12},
    number = {7}
}

@article{Safeguard_Power,
    author = {Perugini, Marco and Gallucci, Marcello and Costantini, Giulio},
    title = {Safeguard Power as a Protection Against Imprecise Power Estimates.},
    journal = {Perspectives on psychological science : a journal of the Association for Psychological Science},
    year = {2014},
    volume = {9},
    number = {3},
    pages = {319-332}
}

@article{Reinagel2023NHacking,
  author  = {Reinagel, Pamela},
  title   = {Is {N-hacking ever OK?} The consequences of collecting more data in pursuit of statistical significance},
  journal = {PLOS Biology},
  year    = {2023},
  volume  = {21},
  number  = {11},
  pages   = {e3002345},
}

@article{rizopoulos2010jm,
  title={{JM: An R package for the joint modelling of longitudinal and time-to-event data}},
  author={Rizopoulos, Dimitris},
  journal={Journal of Statistical Software},
  volume={35},
  number={9},
  pages={1--33},
  year={2010}
}

@article{Selman2024OrdinalRCT,
  author  = {Selman, Chris J. and Lee, Katherine J. and Ferguson, Kristin N. and Whitehead, Clare L. and Manley, Brett J. and Mahar, Robert K.},
  title   = {Statistical analyses of ordinal outcomes in randomised controlled trials: a scoping review},
  journal = {Trials},
  year    = {2024},
  volume  = {25},
  pages   = {241},
}

@article{Simmons2011FalsePositive,
  author  = {Simmons, Joseph P. and Nelson, Leif D. and Simonsohn, Uri},
  title   = {False-positive psychology: undisclosed flexibility in data collection and analysis allows presenting anything as significant},
  journal = {Psychological Science},
  year    = {2011},
  volume  = {22},
  number  = {11},
  pages   = {1359--1366},
}

@article{Sonabend2022Chacking,
  author  = {Sonabend, Raphael and Bender, Andreas and Vollmer, Sebastian},
  title   = {Avoiding {C-hacking} when evaluating survival distribution predictions with discrimination measures},
  journal = {Bioinformatics},
  year    = {2022},
  volume  = {38},
  number  = {17},
  pages   = {4178--4184},
}

@article{Strech2019_3Rs,
  author  = {Strech, Daniel and Dirnagl, Ulrich},
  title   = {3{R}s missing: animal research without scientific value is unethical},
  journal = {BMJ Open Science},
  year    = {2019},
  volume  = {3},
  number  = {1},
  pages   = {bmjos-2018-000048},
  pmid    = {35047678},
  pmcid   = {PMC8647585},
}

@article{White2023,
  author  = {White, Ian R. and Marley-Zagar, Eszter and Morris, Tim P. and Parmar, Mahesh K. B. and Royston, Patrick and Babiker, Abdel G.},
  title   = {artcat: sample-size calculation for an ordered categorical outcome},
  journal = {The Stata Journal},
  year    = {2023},
  volume  = {23},
  number  = {1},
  pages   = {3--23}
}

@article{Zhao2008,
  author  = {Zhao, Y. D. and Rahardja, D. and Mei, Y.},
  title   = {Sample size calculation for the van {Elteren} test adjusting for ties},
  journal = {Journal of Biopharmaceutical Statistics},
  year    = {2008},
  volume  = {18},
  number  = {6},
  pages   = {1112--1119}
}

\clearpage
\appendix

\renewcommand{\thesection}{A\arabic{section}}
\renewcommand{\thesubsection}{\thesection.\arabic{subsection}}

\renewcommand{\theHsection}{appendix.\arabic{section}}
\renewcommand{\theHsubsection}{\theHsection.\arabic{subsection}}

\setcounter{table}{0}
\renewcommand{\thetable}{A\arabic{table}}
\renewcommand{\theHtable}{appendix.\arabic{table}}

\section{Sample size calculation in the fictitious example}
\label{app:sample_size_formulas}
Using the approximate sample size formulas for ordinal data for the two-sample $t$-test \citep{Kieser2020}, the Wilcoxon--Mann--Whitney test \citep{Zhao2008}, and a proportional odds model \citep{White2023}, the required sample sizes per group,  rounded up to the next integer, are calculated as

\begin{align*}
n_{t}
&=
\left\lceil\frac{(1+r)^2}{r}
(z_{1-\alpha/2}+z_{1-\beta})^2
\left(\frac{\sigma}{\delta}\right)^2
+
\frac{z_{1-\alpha/2}^2}{2}\right\rceil
,
\\[6pt]
n_{\text{WMW}}
&=
\left\lceil\frac{(1+r)^2}{r}
\frac{(z_{1-\alpha/2}+z_{1-\beta})^2}{12}
\frac{1-\sum_{i=1}^{M}\bar{p}_i^{3}}
     {(\pi-\tfrac12)^2}\right\rceil
,
\\[6pt]
n_{\text{PO}}
&=
\left\lceil\frac{V_N (z_{1-\alpha/2}+z_{1-\beta})^2}{\theta_A^2}\right\rceil
.
\end{align*}

\noindent Here, $z_q$ denotes the $q$-quantile of the standard normal distribution and $r=n_I/n_C$ the allocation ratio between intervention and control group. For the $t$-test, $\delta=\mu_I - \mu_C$ denotes the assumed difference in group means and $\sigma$ is the average within-group standard deviation defined as $\sigma=\tfrac12(\sigma_C+\sigma_I)$, where for a number of $M$ outcome categories $\sigma_C=\sqrt{\sum_{i=1}^{M}p_{Ci}(i-\mu_C)^2}$ and $\sigma_I=\sqrt{\sum_{i=1}^{M}p_{Ii}(i-\mu_I)^2}$ are the group-specific standard deviations derived from the assumed category probabilities. In the Wilcoxon--Mann--Whitney framework, $\pi=\sum_{i=2}^M(p_{Ii} \sum_{j=1}^{i-1}p_{Cj}) + 0.5  \sum_{i=1}^Mp_{Ci}p_{Ii}$ denotes the relative intervention effect and $\bar p_i = (\frac{rp_{Ci}+p_{Ii}}{1+r})$ the average probability of category $i$. For the proportional odds model, $\theta_A$ denotes the estimated effect in the PO model under the alternative hypothesis and $V_N$ the corresponding variance under the null hypothesis as defined in \cite{White2023}.

\section{Modifications of parameter values of the Joint Model used for data generation}
\label{app:joint_model_parameters}

\begin{table}[!htbp]
\centering
\caption{Originally fitted and modified parameter values of the joint model used for data generation. Two treatment effect scenarios (moderate and large) were considered.}
\label{tab:parameter_modifications}
\begin{tabular*}{0.8\textwidth}{@{\extracolsep{\fill}}l|rr@{\extracolsep{\fill}}}
\toprule
\textbf{Parameter} & \textbf{Fitted} & \textbf{Modified}\\ 
\midrule
Treatment main effect ($\beta_1$) 
& $-0.3848 $ & $-0.05$ \\
Treatment $\times$ time interaction ($\beta_3$) 
& $-0.0397$ & 
\begin{tabular}[t]{@{}l@{}}
moderate: $-0.02$ \\
large: \hfill $-0.04$
\end{tabular} \\
Residual standard deviation ($\sigma$) 
& $0.5470$ & $0.05$ \\
Intercept of survival linear predictor ($\gamma_0$) 
& $-14.0980$ & $-16$ \\
\bottomrule
\end{tabular*}
\end{table}

\clearpage

\section{Additional Results}
\label{app:additional_results}

\begin{table}[!htbp]
\centering
\caption{Number of remaining preliminary datasets per $n_{prelim}$ and effect size with follow-up bias (only methods with calculated sample sizes below the thresholds of $n_{FUB}=50$ or $n_{FUB}=25$ are considered) for the $t$-test, Wilcoxon--Mann--Whitney (WMW) test, linear mixed model (LMM), and sample size minimisation} 
\label{tab:remaining_iterations}
\begin{tabular*}{\textwidth}{@{\extracolsep{\fill}}ll|rrrr|rrrr@{\extracolsep{\fill}}}
\toprule
 &  & \multicolumn{4}{c|}{Moderate effect size} & \multicolumn{4}{c}{Large effect size} \\
 \cline{3-6}\cline{7-10}
 & $n_{prelim}$ & 5 & 10 & 15 & 20 & 5 & 10 & 15 & 20 \\
\midrule
FUB 50 & $t$-test t1 & 7028 & 7572 & 7962 & 8325 & 9437 & 9862 & 9962 & 9993 \\ 
   & $t$-test t2 & 6841 & 7219 & 7635 & 7949 & 9345 & 9831 & 9948 & 9979 \\ 
   & $t$-test t3 & 6718 & 7041 & 7368 & 7674 & 9263 & 9803 & 9938 & 9972 \\ 
   & WMW test t1 & 7000 & 7380 & 7875 & 8136 & 9351 & 9816 & 9941 & 9985 \\ 
   & WMW test t2 & 6835 & 7057 & 7519 & 7716 & 9239 & 9769 & 9936 & 9975 \\ 
   & WMW test t3 & 6659 & 6923 & 7229 & 7457 & 9171 & 9735 & 9906 & 9956 \\ 
   & LMM & 6292 & 6401 & 6667 & 6908 & 9123 & 9719 & 9893 & 9954 \\ 
   & minimisation & 8039 & 8092 & 8397 & 8596 & 9648 & 9914 & 9977 & 9995 \\  
   \midrule
   & & & & & & & & &\\
FUB 25 & $t$-test t1 & 5611 & 5646 & 5819 & 5916 & 8821 & 9528 & 9779 & 9886 \\ 
   & $t$-test t2 & 5342 & 5289 & 5369 & 5390 & 8681 & 9398 & 9727 & 9848 \\ 
   & $t$-test t3 & 5205 & 5090 & 5065 & 5025 & 8576 & 9328 & 9661 & 9823 \\ 
   & WMW test t1 & 5309 & 5497 & 5455 & 5374 & 8535 & 9429 & 9711 & 9821 \\ 
   & WMW test t2 & 5113 & 5128 & 4953 & 4886 & 8406 & 9283 & 9598 & 9762 \\ 
   & WMW test t3 & 5008 & 4883 & 4677 & 4521 & 8326 & 9192 & 9503 & 9721 \\ 
   & LMM & 4788 & 4443 & 4264 & 4063 & 8384 & 9164 & 9523 & 9731 \\ 
   & minimisation & 6486 & 6325 & 6300 & 6279 & 9139 & 9657 & 9839 & 9912 \\ 
   \bottomrule
  \end{tabular*}
\end{table}

\begin{table}[!htbp]
\centering
\caption{Median estimated sample size per group per $n_{prelim}$ and effect size without follow-up bias for the $t$-test, Wilcoxon--Mann--Whitney (WMW) test, linear mixed model (LMM), and sample size minimisation. The columns ``true'' show the respective true sample sizes (estimated with the parameters from the large dataset) to reach a power of 0.8} 
\label{tab:median_sample_size}
\begin{tabular*}{\textwidth}{@{\extracolsep{\fill}}l|rrrrr|rrrrr@{\extracolsep{\fill}}}
\toprule
 & \multicolumn{5}{c|}{Moderate effect size} & \multicolumn{5}{c}{Large effect size} \\
 \cline{2-6}\cline{7-11}
$n_{prelim}$ & 5 & 10 & 15 & 20 & true & 5 & 10 & 15 & 20 & true \\
\midrule
$t$-test t1 & 20 & 22 & 22 & 22 & 22 & 7 & 8 & 8 & 8 & 8 \\ 
  $t$-test t2 & 22 & 24 & 24 & 24 & 24 & 8 & 8 & 8 & 8 & 8 \\ 
  $t$-test t3 & 24 & 25 & 25 & 25 & 26 & 8 & 8 & 8 & 8 & 8 \\ 
  WMW test t1 & 20 & 23 & 24 & 24 & 25 & 10 & 10 & 10 & 10 & 10 \\ 
  WMW test t2 & 21 & 25 & 26 & 26 & 27 & 10 & 10 & 10 & 10 & 11 \\ 
  WMW test t3 & 25 & 27 & 28 & 28 & 29 & 10 & 10 & 10 & 10 & 11 \\ 
  LMM & 28 & 31 & 31 & 31 & 32 & 8 & 8 & 8 & 8 & 8 \\ 
  minimisation & 15 & 18 & 20 & 20 &  & 7 & 7 & 7 & 8 &  \\ 
\bottomrule   
\end{tabular*}
\end{table}

\begin{table}[!htbp]
\centering
\caption{Median estimated sample size per group per $n_{prelim}$ and effect size with follow-up bias (only methods with calculated sample sizes below the threshold of $n_{FUB}=50$ are considered) for the $t$-test, Wilcoxon--Mann--Whitney (WMW) test, linear mixed model (LMM), and sample size minimisation. The columns ``true'' show the respective true sample sizes (estimated with the parameters from the large dataset) to reach a power of 0.8} 
\label{tab:median_sample_size_followup50}
\begin{tabular*}{\textwidth}{@{\extracolsep{\fill}}l|rrrrr|rrrrr@{\extracolsep{\fill}}}
 \toprule
 & \multicolumn{5}{c|}{Moderate effect size} & \multicolumn{5}{c}{Large effect size} \\
 \cline{2-6}\cline{7-11}
$n_{prelim}$ & 5 & 10 & 15 & 20 & true & 5 & 10 & 15 & 20 & true \\
\midrule
$t$-test t1 & 13 & 16 & 18 & 19 & 22 & 7 & 8 & 8 & 8 & 8 \\ 
  $t$-test t2 & 13 & 17 & 18 & 20 & 24 & 7 & 8 & 8 & 8 & 8 \\ 
  $t$-test t3 & 14 & 17 & 19 & 21 & 26 & 7 & 8 & 8 & 8 & 8 \\ 
  WMW test t1 & 15 & 17 & 19 & 21 & 25 & 8 & 10 & 10 & 10 & 10 \\ 
  WMW test t2 & 15 & 18 & 21 & 22 & 27 & 10 & 10 & 10 & 10 & 11 \\ 
  WMW test t3 & 15 & 18 & 21 & 22 & 29 & 10 & 10 & 10 & 10 & 11 \\ 
  LMM & 14 & 18 & 20 & 22 & 32 & 7 & 8 & 8 & 8 & 8 \\ 
  minimisation & 12 & 15 & 17 & 18 &  & 7 & 7 & 7 & 8 &  \\ 
\bottomrule   
\end{tabular*}
\end{table}

\begin{table}[!htbp]
\centering
\caption{Median estimated sample size per group per $n_{prelim}$ and effect size with follow-up bias (only methods with calculated sample sizes below the threshold of $n_{FUB}=25$ are considered) for the $t$-test, Wilcoxon--Mann--Whitney (WMW) test, linear mixed model (LMM), and sample size minimisation. The columns ``true'' show the respective true sample sizes (estimated with the parameters from the large dataset) to reach a power of 0.8} 
\label{tab:median_sample_size_followup25}
\begin{tabular*}{\textwidth}{@{\extracolsep{\fill}}l|rrrrr|rrrrr@{\extracolsep{\fill}}}
  \toprule
 & \multicolumn{5}{c|}{Moderate effect size} & \multicolumn{5}{c}{Large effect size} \\
 \cline{2-6}\cline{7-11}
$n_{prelim}$ & 5 & 10 & 15 & 20 & true & 5 & 10 & 15 & 20 & true \\
\midrule
$t$-test t1 & 10 & 13 & 14 & 15 & 22 & 7 & 8 & 8 & 8 & 8 \\ 
  $t$-test t2 & 11 & 13 & 15 & 16 & 24 & 7 & 8 & 8 & 8 & 8 \\ 
  $t$-test t3 & 11 & 13 & 15 & 16 & 26 & 7 & 8 & 8 & 8 & 8 \\ 
  WMW test t1 & 12 & 15 & 16 & 17 & 25 & 8 & 10 & 10 & 10 & 10 \\ 
  WMW test t2 & 12 & 15 & 16 & 17 & 27 & 8 & 10 & 10 & 10 & 11 \\ 
  WMW test t3 & 12 & 15 & 16 & 17 & 29 & 8 & 10 & 10 & 10 & 11 \\ 
  LMM & 10 & 14 & 15 & 16 & 32 & 6 & 7 & 8 & 8 & 8 \\ 
  minimisation & 10 & 13 & 14 & 15 &  & 6 & 7 & 7 & 8 &  \\ 
\bottomrule   
\end{tabular*}
\end{table}

\begin{table}[!htbp]
\centering
\caption{Fraction of underpowered iterations (power $<$ 0.8) per $n_{prelim}$ and effect size with follow-up bias (only methods with calculated sample sizes below the threshold of $n_{FUB}=50$ are considered) for the $t$-test, Wilcoxon--Mann--Whitney (WMW) test, linear mixed model (LMM), and sample size minimisation} 
\label{tab:prob_below0.8_followup50}
\begin{tabular*}{\textwidth}{@{\extracolsep{\fill}}l|rrrr|rrrr@{\extracolsep{\fill}}}
  \toprule
 & \multicolumn{4}{c|}{Moderate effect size} & \multicolumn{4}{c}{Large effect size} \\
 \cline{2-5}\cline{6-9}
$n_{prelim}$ & 5 & 10 & 15 & 20 & 5 & 10 & 15 & 20 \\
\midrule
$t$-test t1 & 0.731 & 0.660 & 0.624 & 0.592 & 0.532 & 0.475 & 0.465 & 0.451 \\ 
  $t$-test t2 & 0.749 & 0.692 & 0.653 & 0.622 & 0.509 & 0.454 & 0.434 & 0.408 \\ 
  $t$-test t3 & 0.775 & 0.723 & 0.687 & 0.655 & 0.507 & 0.435 & 0.408 & 0.383 \\ 
  WMW test t1 & 0.758 & 0.701 & 0.644 & 0.628 & 0.504 & 0.471 & 0.453 & 0.434 \\ 
  WMW test t2 & 0.748 & 0.728 & 0.684 & 0.661 & 0.609 & 0.547 & 0.558 & 0.535 \\ 
  WMW test t3 & 0.851 & 0.759 & 0.713 & 0.686 & 0.595 & 0.536 & 0.536 & 0.510 \\ 
  LMM & 0.840 & 0.796 & 0.765 & 0.741 & 0.535 & 0.472 & 0.451 & 0.430 \\ 
  minimisation & 0.798 & 0.747 & 0.698 & 0.668 & 0.610 & 0.536 & 0.515 & 0.494 \\ 
\bottomrule   
\end{tabular*}
\end{table}

\begin{table}[!htbp]
\centering
\caption{Fraction of underpowered iterations (power $<$ 0.8) per $n_{prelim}$ and effect size with follow-up bias (only methods with calculated sample sizes below the threshold of $n_{FUB}=25$ are considered) for the $t$-test, Wilcoxon--Mann--Whitney (WMW) test, linear mixed model (LMM), and sample size minimisation} 
\label{tab:prob_below0.8_followup25}
\begin{tabular*}{\textwidth}{@{\extracolsep{\fill}}l|rrrr|rrrr@{\extracolsep{\fill}}}
  \toprule
 & \multicolumn{4}{c|}{Moderate effect size} & \multicolumn{4}{c}{Large effect size} \\
 \cline{2-5}\cline{6-9}
$n_{prelim}$ & 5 & 10 & 15 & 20 & 5 & 10 & 15 & 20 \\
\midrule
$t$-test t1 & 0.915 & 0.885 & 0.853 & 0.834 & 0.569 & 0.492 & 0.474 & 0.456 \\ 
  $t$-test t2 & 0.959 & 0.944 & 0.929 & 0.917 & 0.548 & 0.475 & 0.443 & 0.413 \\ 
  $t$-test t3 & 1.000 & 1.000 & 1.000 & 1.000 & 0.547 & 0.457 & 0.420 & 0.389 \\ 
  WMW test t1 & 1.000 & 0.941 & 0.930 & 0.951 & 0.552 & 0.490 & 0.464 & 0.441 \\ 
  WMW test t2 & 1.000 & 1.000 & 1.000 & 1.000 & 0.669 & 0.575 & 0.578 & 0.547 \\ 
  WMW test t3 & 1.000 & 1.000 & 1.000 & 1.000 & 0.656 & 0.567 & 0.559 & 0.522 \\ 
  LMM & 1.000 & 1.000 & 1.000 & 1.000 & 0.582 & 0.500 & 0.469 & 0.440 \\ 
  minimisation & 0.979 & 0.951 & 0.925 & 0.910 & 0.644 & 0.550 & 0.523 & 0.498 \\ 
\bottomrule
  \end{tabular*}
\end{table}

\FloatBarrier

\begin{table}[!htbp]
\centering
\small
\caption{Percentage of preliminary datasets for which each method yielded the smallest estimated sample size used for sample size minimisation per $n_{prelim}$ and effect size without and with follow-up bias (only methods with calculated sample sizes below the thresholds of $n_{FUB}=50$ or $n_{FUB}=25$ are considered) for the $t$-test, Wilcoxon--Mann--Whitney (WMW) test and linear mixed model (LMM). In case of ties, each method with the minimal sample size received a weight of $\frac1k$, where $k$ denotes the number of tied methods.} 
\label{tab:minfreq_all}
\begin{tabular*}{\textwidth}{@{\extracolsep{\fill}}ll|rrrr|rrrr@{\extracolsep{\fill}}}
   \toprule
 &  &  \multicolumn{4}{c|}{Moderate effect size} & \multicolumn{4}{c}{Large effect size} \\
 \cline{3-6}\cline{7-10}
 & $n_{prelim}$ & 5 & 10 & 15 & 20 & 5 & 10 & 15 & 20 \\
\midrule
No FUB & $t$-test t1 & 25.6 & 35.7 & 42.9 & 49.5 & 24.9 & 32.7 & 36.7 & 39.9 \\ 
   & $t$-test t2 & 13.2 & 16.4 & 17.1 & 17.3 & 13.5 & 16.9 & 18.1 & 18.7 \\ 
   & $t$-test t3 & 10.7 & 11.4 & 10.8 & 10.1 & 10.8 & 12.1 & 12.1 & 12.1 \\ 
   & WMW test t1 & 15.4 & 15.5 & 14.4 & 12.9 & 7.5 & 4.1 & 2.3 & 1.3 \\ 
   & WMW test t2 & 11.4 & 8.5 & 7.2 & 5.1 & 5.3 & 2.3 & 1.3 & 0.5 \\ 
   & WMW test t3 & 10.2 & 6.6 & 4.6 & 3.4 & 4.7 & 1.8 & 0.7 & 0.3 \\ 
   & LMM & 13.6 & 6.0 & 3.1 & 1.6 & 33.2 & 30.2 & 28.8 & 27.1 \\ 
\midrule
FUB 50 & $t$-test t1 & 27.1 & 37.3 & 44.6 & 51.4 & 25.0 & 32.6 & 36.7 & 39.9 \\ 
   & $t$-test t2 & 14.3 & 17.9 & 18.2 & 17.9 & 13.4 & 16.8 & 18.1 & 18.7 \\ 
   & $t$-test t3 & 11.2 & 12.5 & 11.6 & 10.6 & 10.8 & 12.2 & 12.1 & 12.2 \\ 
   & WMW test t1 & 14.5 & 13.9 & 12.5 & 11.4 & 7.2 & 3.9 & 2.2 & 1.3 \\ 
   & WMW test t2 & 10.5 & 7.4 & 6.3 & 4.4 & 4.9 & 2.3 & 1.2 & 0.5 \\ 
   & WMW test t3 & 8.8 & 5.6 & 4.2 & 2.8 & 4.4 & 1.8 & 0.7 & 0.3 \\ 
   & LMM & 13.6 & 5.4 & 2.7 & 1.5 & 34.2 & 30.4 & 28.8 & 27.1 \\ 
   \midrule
FUB 25 & $t$-test t1 & 28.8 & 38.5 & 47.0 & 53.8 & 24.8 & 32.6 & 36.6 & 39.8 \\ 
   & $t$-test t2 & 14.9 & 19.2 & 19.6 & 19.5 & 13.6 & 16.8 & 18.1 & 18.7 \\ 
   & $t$-test t3 & 11.8 & 13.4 & 12.9 & 11.6 & 11.0 & 12.2 & 12.2 & 12.2 \\ 
   & WMW test t1 & 12.4 & 11.7 & 9.2 & 7.8 & 6.3 & 3.5 & 2.1 & 1.1 \\ 
   & WMW test t2 & 8.8 & 6.0 & 4.7 & 3.2 & 4.4 & 2.1 & 1.1 & 0.5 \\ 
   & WMW test t3 & 7.6 & 4.4 & 3.1 & 2.2 & 3.9 & 1.6 & 0.7 & 0.3 \\ 
   & LMM & 15.7 & 6.7 & 3.5 & 2.0 & 36.0 & 31.2 & 29.2 & 27.4 \\ 
\bottomrule  
  \end{tabular*}
\end{table}
\end{document}